\documentclass{article}
\usepackage{bigsky2007}
\usepackage{graphicx}
\frompage{000} \topage{000}                                              %|
\def\rightmark{Resonance Production in Jets }
\def\leftmark{Christina Markert et al.}

\title{Resonance Production in Jets}
\authors{
{Christina Markert$^1$ for the STAR Collaboration %
}\\[2.812mm]
{\normalsize
\hspace*{-8pt}$^1$ University of Texas, Austin, Texas 78712, USA\\[0.2ex]
}}

\abstract{Hadronic resonances with short life times and strong
coupling to the dense medium may exhibit mass shifts and width
broadening as signatures of chiral symmetry restoration at the phase
transition between hadronic and partonic matter. Resonances with
different lifetimes are also used to extract information about the
time evolution and temperature of the expanding hadronic medium. In
order to collect information about the early stage (at the phase
transition) of a heavy-ion collision, resonances and decay particles
which are unaffected by the hadronic medium have to be used. We
explore a possible new technique to extract signals from the early
stage through the selection of resonances from jets. A first attempt
of this analysis, using the reconstructed $\phi$(1020) from 200 GeV
Au+Au collisions in STAR, is presented.}

\keyword{resonance, jets, lifetime, strange, freeze-out, medium}
\PACS{25.75.Dw}
\begin{document}

\maketitle
\setcounter{page}{1}

\section{Introduction}\label{intro}

Yields of resonances measured via their hadronic decays are
sensitive to the hadronic lifetime of the nuclear medium in a
heavy-ion reaction. Together with the pion HBT lifetime measurement
($\Delta\tau=5-12$~fm/c) ~\cite{nig05}, which determines the time
from the beginning of the collision to kinetic freeze-out, we,
extract a partonic lifetime, under the assumption that the chemical
freeze-out occurs at hadronization \cite{resostar}. The extended
hadronic medium which exists between chemical and kinetic freeze-out
in heavy-ion reactions may change the yield and spectra of
resonances due to the re-scattering of the decay particles and
possible regeneration of resonances. The resonance over
non-resonance ratios from STAR \cite{resostar} can be described with
a microscopic transport model (UrQMD) which assumes a lifetime
between chemical and kinetic freeze-out of $\Delta \tau = 10
 \pm 3 $~fm/c \cite{urqmd}. Alternatively, suppression of the
$\Lambda$(1520) and K(892) yields in Au+Au collisions combined with
a thermal model with an additional re-scattering phase
\cite{tor01,raf01,raf02,mar02} were used to estimate a hadronic
lifetime of $\Delta \tau > 4$~fm/c \cite{resostar}. \\
To study chiral symmetry restoration in terms of mass shifts and
width broadening, resonance decays from the early stage of the
medium need to be extracted. This can be done by reconstructing
leptonic decay channels because leptons are less likely to
re-scatter in the hadronic medium. However, regenerated resonances
from the late hadronic phase feed into this signal. According to
UrQMD calculations, the re-scattering and regeneration processes
only change the low momentum region (p$_{\rm T}$ = 0-2 GeV/c) of the
resonance spectra. Therefore high momentum resonances and their
decay particles are less likely to be affected by the hadronic
medium. In order to test the impact of the partonic medium on
resonance properties we need to reconstruct high momentum resonances
which are produced early. This might be possible through the
selection of resonances from the away-side jet of a triggered di-jet
event.

%which is the away-side region
%($\Delta\phi= \pi$) from angular correlation of a trigger and
%associated resonance particle.

%trigger/ event =0.12
%phi(1020)/event = 0.05

\section{Jet Resonance Correlations}

\begin{figure}[h!]
\centering
\includegraphics[width=0.75\textwidth]{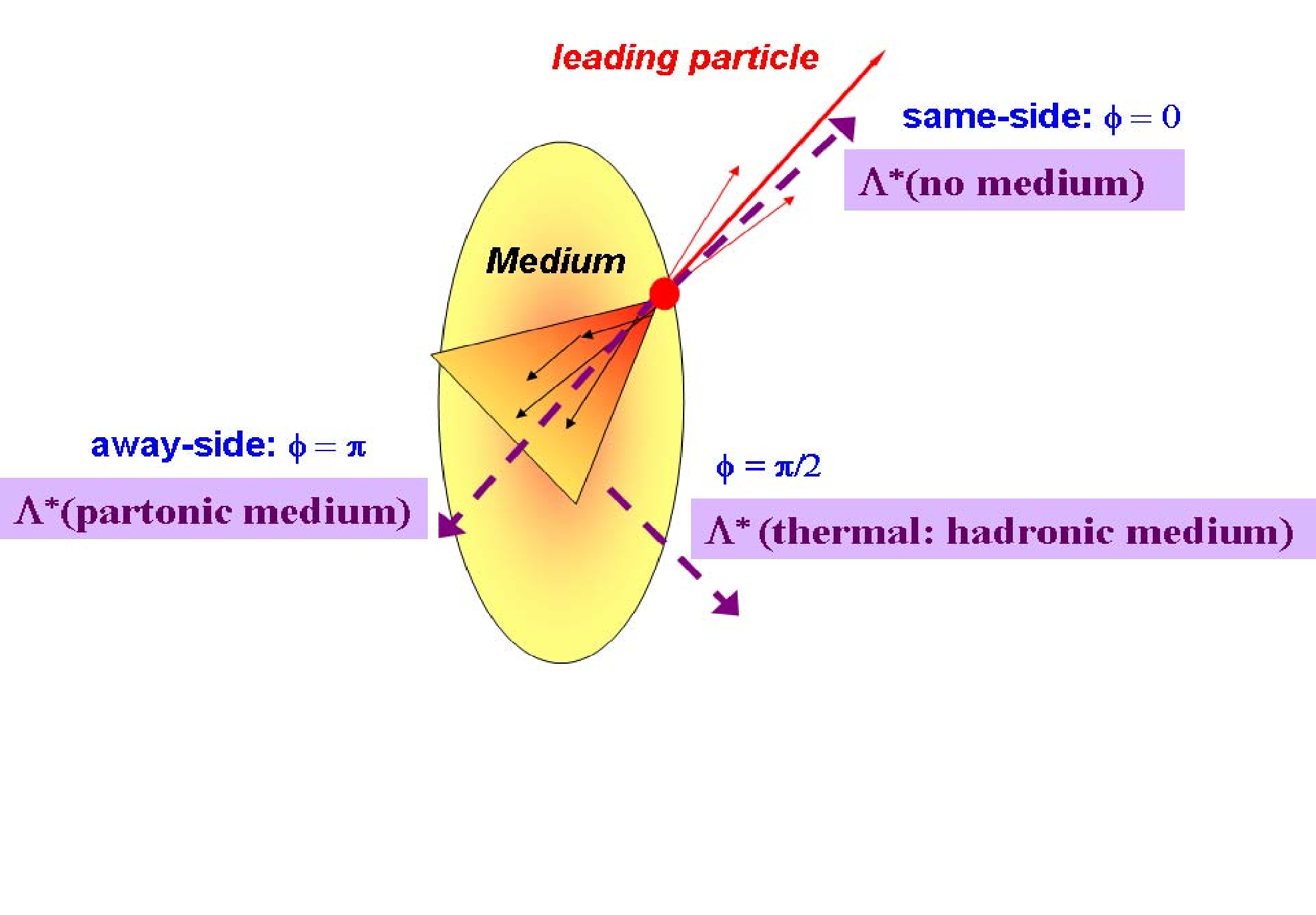}
%\vspace{-10mm}
 \caption{Sketch of jet fragmentation into
resonances ($\Lambda$*, $\phi$(1020),...) in the medium created in a
heavy-ion collision. Same-side correlations of resonances are not
affected by the medium, whereas the away-side high p$_{\rm T}$
resonance might be affected by the early (chiral restored) medium.
Thermal resonances, which are affected by the late hadronic medium
are at $\pi$/2 with respect to the trigger particle.}
 \label{jetresosketch}
\end{figure}

A leading trigger-particle correlation analysis requires a high
momentum trigger particle to identify the jet axis and the jet side
which is less affected by the medium (same-side). Therefore, the
away-side ($\Delta\phi=\pi$) correlations will measure the medium
modified jet. High momentum resonances from the away-side jet are
identified via the angle with respect to the jet axis or leading
particle (see Figure~\ref{jetresosketch}). A high transverse
momentum resonance in the away-side jet cone is likely to be
produced early, which, depending on its formation time, can interact
with the early partonic medium, but leaves the medium fast enough to
not exhibit any interaction in the late hadronic phase. The
formation time of a resonance in the string fragmentation process
depends on the momentum fraction $z$ carried by the resonance. In
addition there is a parton and resonance mass dependence which leads
to shorter formation times for heavy resonances. Two approaches were
proposed recently: firstly, a study based on the string
fragmentation implementation in PYTHIA \cite{falter}; secondly, a
quantum mechanical treatment of heavy meson formation in heavy-ion
collisions \cite{vitev}. Both cases demonstrate that the probability
of high momentum heavy hadron (or
resonance) formation in the partonic medium is finite.\\
Quantitative studies of resonance properties such as yield, mass,
width, and branching ratio as a function of resonance momentum,
emission angle, jet energy, and jet tag, will directly address the
question of chiral symmetry restoration. The low momentum resonances
produced at an angle of $\Delta \phi = 1/2 \pi$ or $3/2 \pi$ with
respect to the jet axis or leading particle are identified as late
produced thermal resonances from the bulk matter of the collision.
They predominantly interact in the late hadronic medium. Therefore
their masses and widths are expected to be in agreement with vacuum
conditions unless phase space effects from late regeneration change
the shape of the invariant mass signal. The high p$_{\rm T}$
same-side jet resonances are also expected to exhibit the vacuum
width and mass because the initial quark is expected to fragment
outside of the medium (surface bias effect). Therefore a
differential measurement of high momentum resonances as a function
of the angle to the jet axis might have a built-in reference system
and may not require comparison to measurements in proton+proton
collisions, which are very statistics limited for the rare hadronic
resonance channels.

\section{Resonance Correlations in STAR}

We recently initiated a first attempt to study the high momentum
charged hadron-$\phi$(1020) resonance correlation using the STAR
detector. We are using the $\phi$(1020) as the associated particle
since its reconstructed mass spectrum has the largest significance
of all short lived resonances in STAR. However the lifetime of the
$\phi$(1020) is about 45 fm/c, which means that the majority will
decay outside of the medium. The number of Au+Au events analyzed
(4.5M 0-20\% most central) is not sufficient to place an effective
high momentum cut on the $\phi$(1020) spectrum and therefore we are
not sensitive to mass shifts or width broadenings. The momentum of
the kaon candidates for the $\phi$(1020) reconstruction is
restricted to p$_{\rm T}$ [0.2-1.0] GeV/c in order to achieve clean
pid in the TPC. The mean transverse momentum of the $\phi$(1020) is
$\langle p_{\rm T} \rangle \sim 0.9$ GeV/c. Therefore, most of the
reconstructed $\phi$(1020) resonances are from the thermal medium
rather than from an early fragmenting jet. In the future we need to
select the higher momentum $\phi$(1020)s with better significance by
using the additional Time of Flight (TOF) detector ~\cite{tof} which
allows us to identify kaons up to p = 1.5-2 GeV/c. We will show in
these proceedings the analysis techniques to study resonances from
jets and discuss the potential results in general terms.

%trigger/ event =0.12
%phi(1020)/event = 0.05

\section{Angle dependent Invariant Mass Distribution of $\phi$(1020)}

\begin{figure}[h!]
\centering
\includegraphics[width=0.80\textwidth]{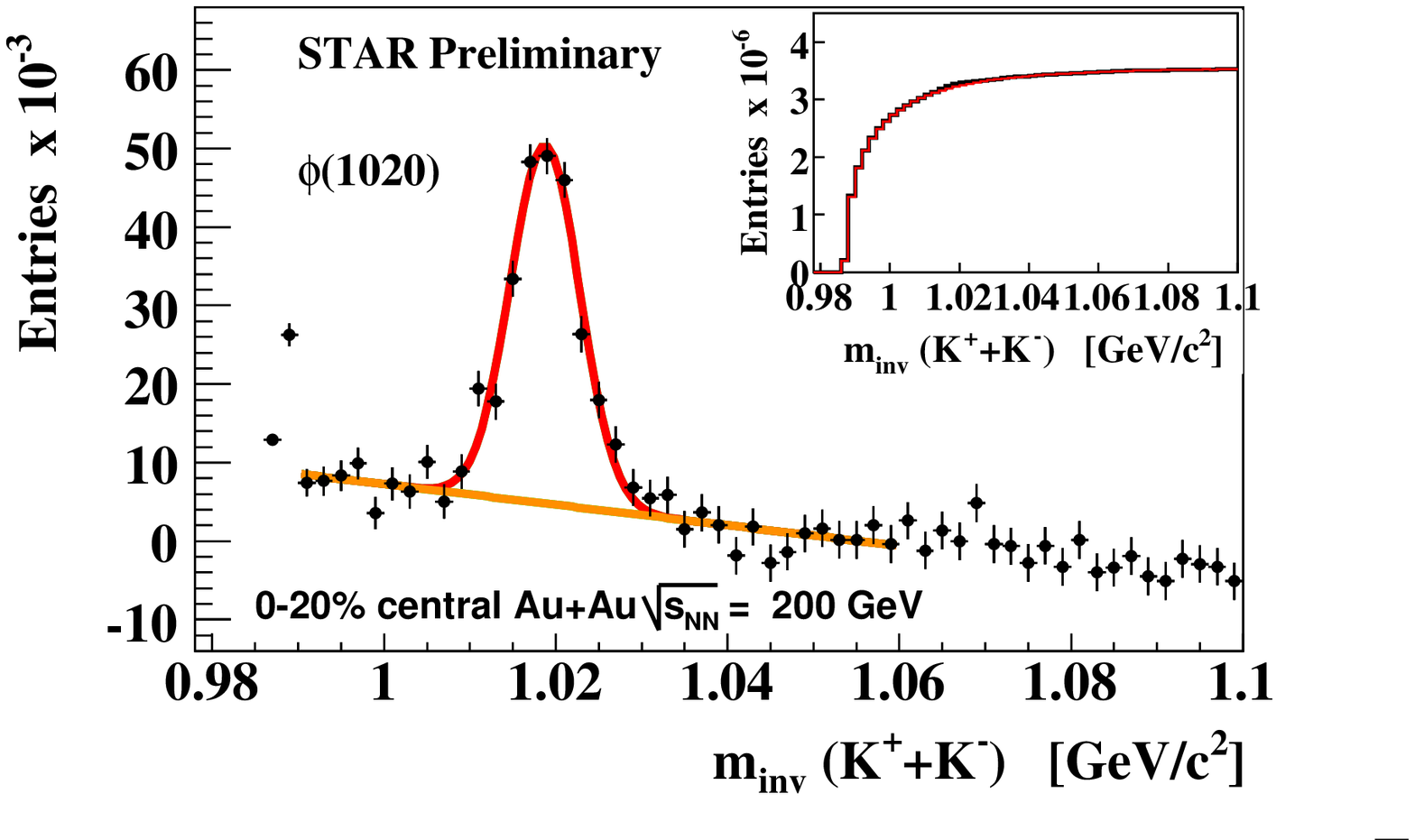}
\caption{Invariant mass distributions before and after mixed-event
background subtraction for $\phi$(1020) from 4.5 M 0-20\% most
central 200~GeV Au+Au events including at least one hadron with
p$_{T}$~$>$~4~GeV/c.}
\label{invall}
\end{figure}

Hadrons with p$_{T}$~$>$~4~GeV/c are selected as jet trigger
particles (p$_{\rm T}^{\rm trig}$/event = 0.12) and correlated with
$\phi$(1020)s ($\langle p_{\rm T} \rangle \sim 0.9$ GeV/c) as the
associated particles. Figure~\ref{invall} shows the $\phi$(1020)
invariant mass distributions before and after mixed-event background
subtraction for 4.5 Million 0-20\% most central Au+Au collisions
containing at least one charged hadron with p$_{T}$~$>$~4~GeV/c. The
number of entries in the $\phi$(1020) signal peak is about 230,000
($\phi$(1020)/event = 0.05). The derived $\phi$(1020) mass is m =
1.0188 $\pm$ 0.0002 GeV/c$^{2}$ and the width is $\Gamma$ = 4.0
$\pm$ 0.2 MeV/c$^{2}$ from a Gaussian combined with a linear fit,
which are in agreement with the PDG value, folded with the
detector momentum resolution and the energy loss in the detectors. \\

\begin{figure}[h!]
\centering
\includegraphics[width=0.48\textwidth]{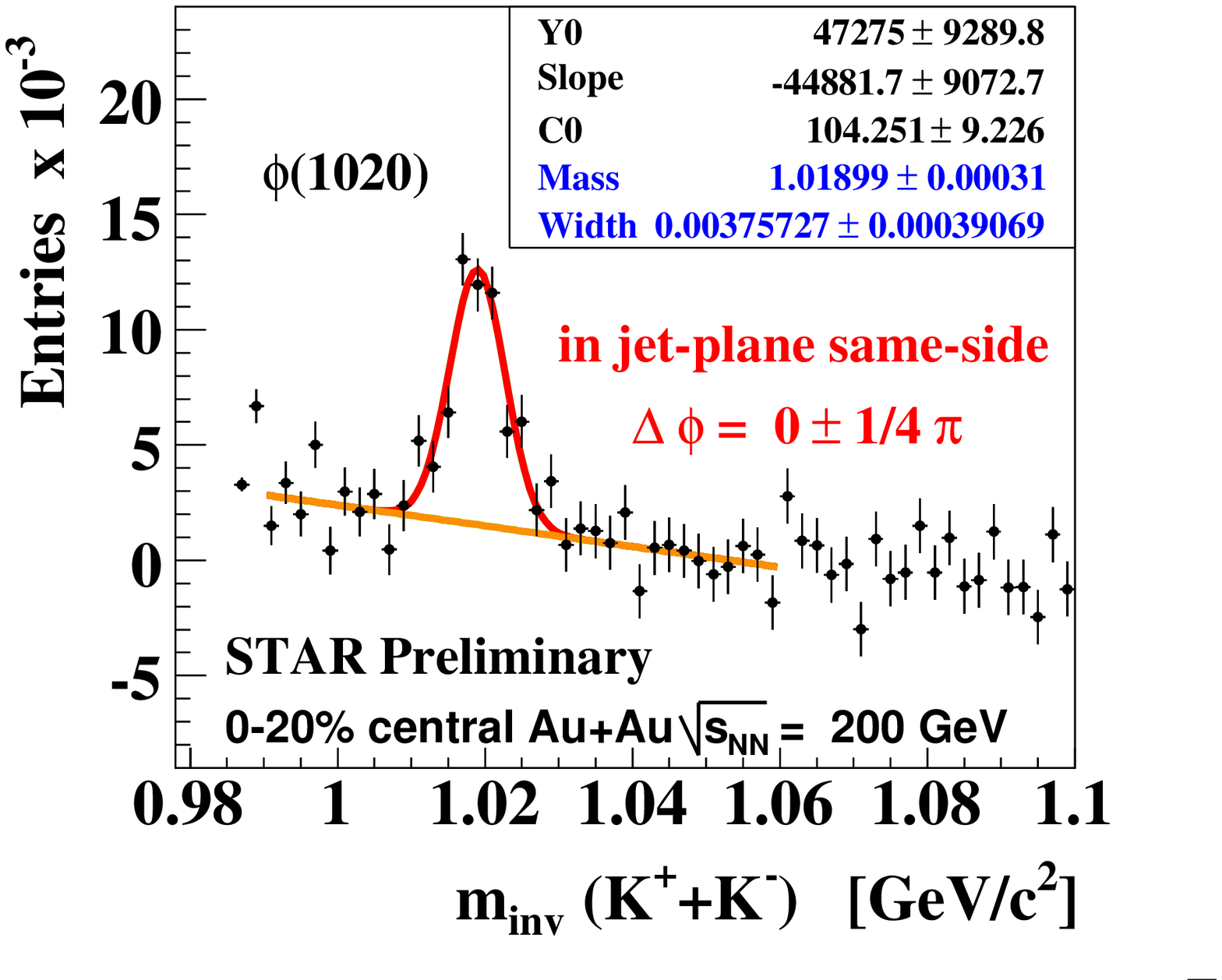}
\includegraphics[width=0.48\textwidth]{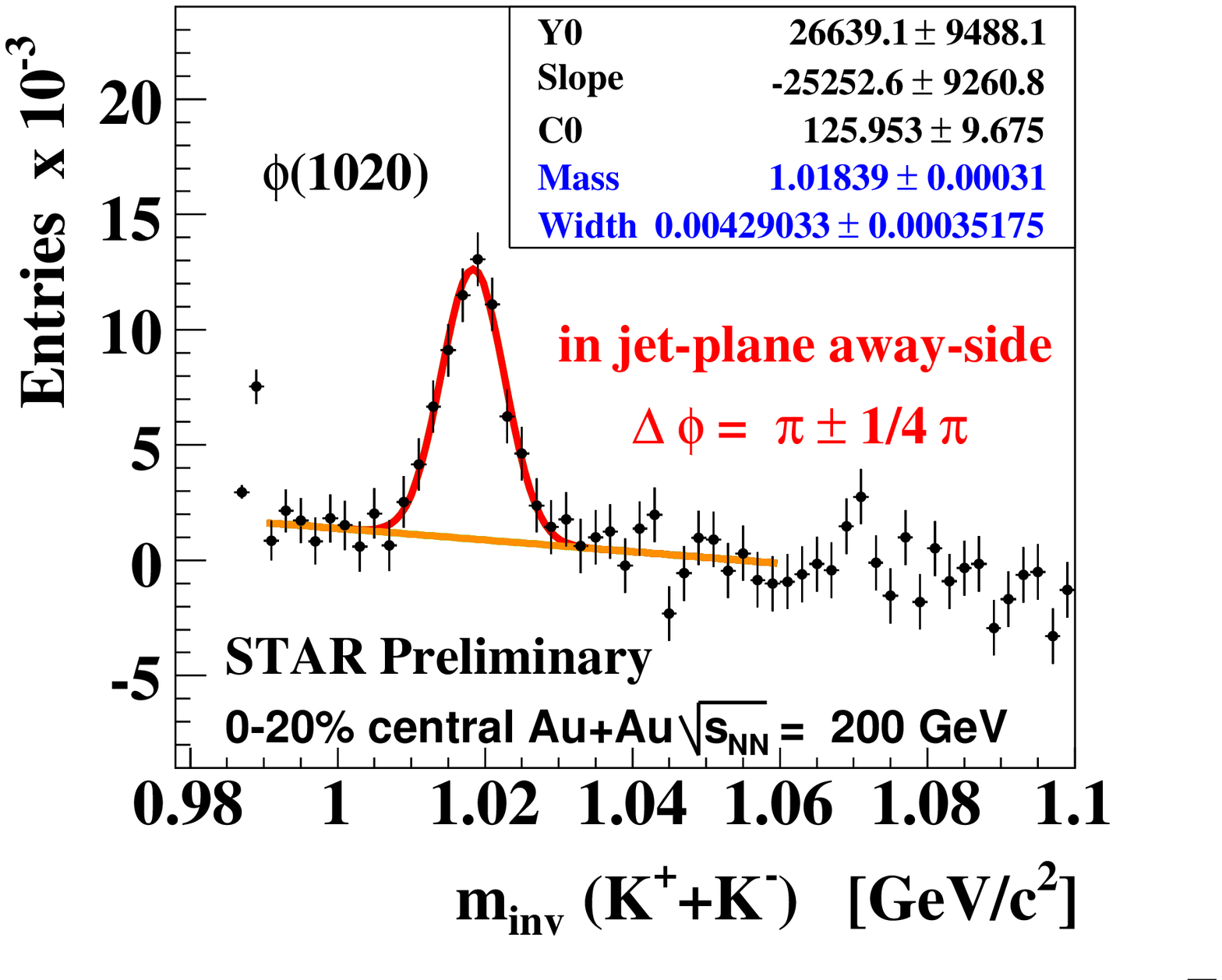}
\caption{$\phi$(1020) invariant mass distributions after mixed-event
background subtraction for their same-side (left) and away-side
(right) angular correlation with respect to the trigger hadron of
p$_{T}$~$>$~4~GeV/c.} \label{invinplane}
\end{figure}

\begin{figure}[h!]
\centering
\includegraphics[width=0.48\textwidth]{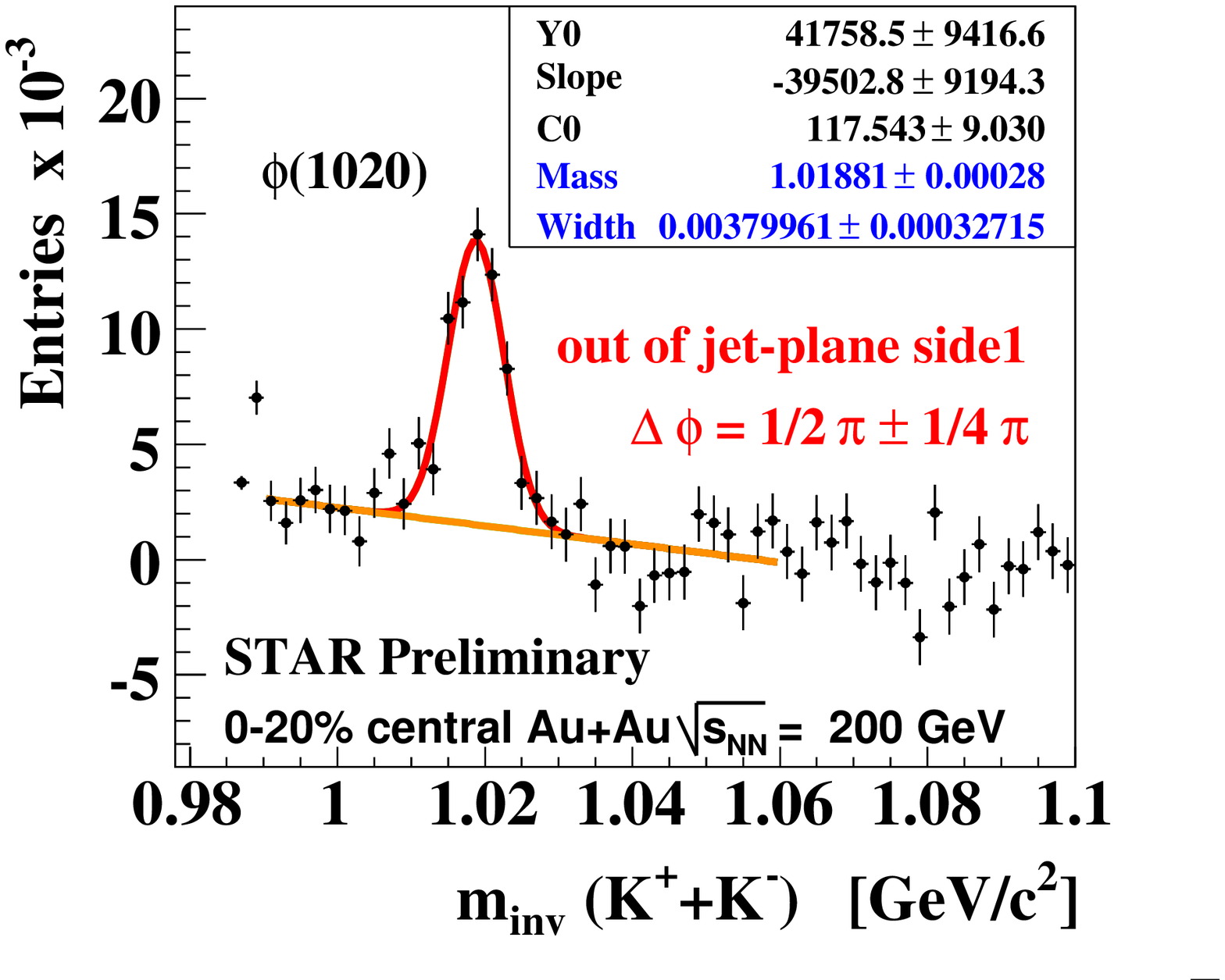}
\includegraphics[width=0.48\textwidth]{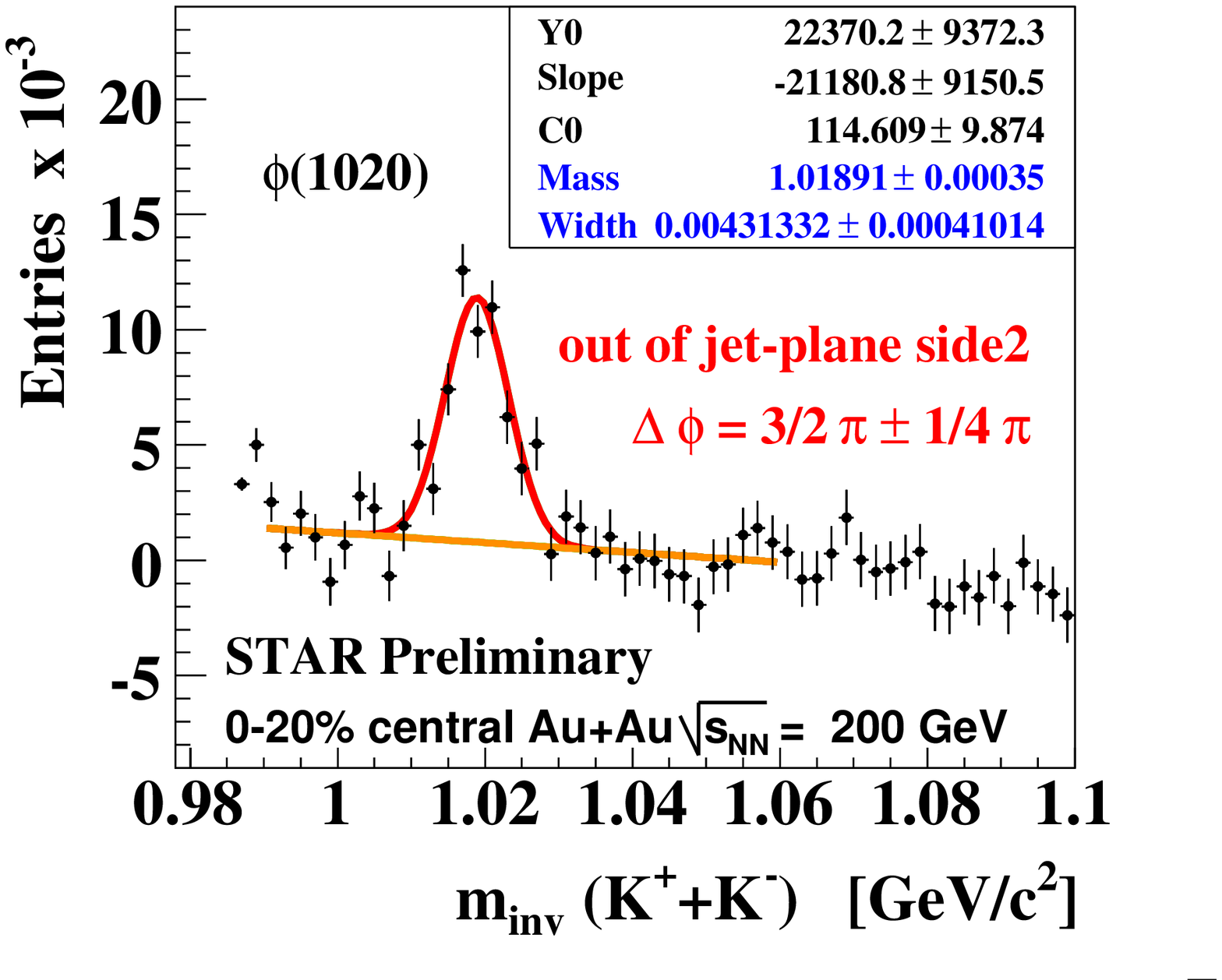}
\caption{$\phi$(1020) invariant mass distributions after mixed-event
background subtraction for the out of jet plane angular correlation
$\Delta$$\phi$ = $[1/4 \pi, +3/4 \pi]$ (left) and $\Delta$$\phi$ =
$[5/4 \pi, +7/4 \pi]$ (right) with respect to the trigger hadron of
p$_{T}$~$>$~4~GeV/c.} \label{invoutplane}
\end{figure}

We divide the signal into 4 categories according to the $\phi$(1020)
angular orientation with respect to the leading (jet) particle. Two
quadrants in the jet plane: 1. same-side $\Delta$$\phi$ = $[-1/4
\pi, +1/4 \pi]$: 2. away-side $\Delta$$\phi$ = $[+3/4 \pi, +5/4
\pi]$ and two quadrants out of the jet plane: 3. with $\Delta$$\phi$
= $[+1/4 \pi, +3/4 \pi]$ and 4. $\Delta$$\phi$ = $+[5/4 \pi, +7/4
\pi]$. Figure~\ref{invinplane} shows the $\phi$(1020) invariant mass
distributions after mixed-event background subtraction for the
same-side (left) and away-side (right) angular correlations with
respect to the trigger hadron. Figure~\ref{invoutplane} shows the
$\phi$(1020) invariant mass signal for the out of jet plane angular
correlations with respect to the trigger hadron. The masses and
widths of the $\phi$(1020) signals for the different angular
selections are in agreement with the PDG value, folded with the
detector momentum resolution. The $\phi$(1020) yields in the mass
region of 1019.5 $\pm$ 7 MeV/c$^{2}$ for the four angle ranges are
shown in table~\ref{yield} with the statistical errors. The
systematical error due to the normalization of the background, the
invariant range and the linear fit is on the order of 10\%. The
yield of the signal on the away-side is 26\%$\pm$19\% higher than on
the same-side, which means that there is a trend of a larger
resonance production in the away-side $\Delta \phi$ correlation
compared to the same-side, which might be due to energy conservation
(trigger bias). This would mean that it is less likely to produce a
massive resonance if the high momentum particle in a jet takes a
large fraction of the energy. From Figure \ref{resocorrstarreso} we
derive that the splitting of $\Delta \phi$ into 4 equal parts might
not contain the full jet on the away-side. The size of the same- and
away-side jet in $\Delta \phi$ has to be studied in more detail.

\begin{table}[h!]
\begin{tabular}{ccc}
correlation bin & angle  & yield\\  \hline
all & $[0, 2 \pi]$ &   231085 $\pm$ 4771  $\pm$  10\%\\
in jet-plane same-side & $[-1/4 \pi, +1/4 \pi]$  & 51385 $\pm$ 2369  $\pm$ 10\% \\
in jet-plane away-side & $[+3/4 \pi, +5/4 \pi]$  &  64498 $\pm$ 2400  $\pm$ 10\%\\
out of jet-plane side 1 & $[+1/4 \pi, +3/4 \pi]$   & 61043 $\pm$ 2394 $\pm$ 10\%\\
out of jet-plane side 2& $[+5/4 \pi, +7/4 \pi]$   &  54893  $\pm$ 2378  $\pm$ 10\% \\
\end{tabular}
\centering \caption{$\phi$(1020) yields in the mass region of 1019.5
$\pm$ 7 MeV/c$^{2}$ for the four angle ranges. } \label{yield}
\end{table}

% inv mass +- 8 MeV
%same side = 56525 +- 2800
%away side = 64398+- 2800
%out of plane 1 : 60128 +- 2800
%out of plane 2 : 60022 +- 2800

% inv mass 1019.5 +- 0.007
% all: 231085 +- 4771
% in plane same: 51385 +- 2369
% in plane away: 64498 +- 2400
% out plane close: 61043 +- 2394
% out plane far: 54893 +- 2378

\section{Charged hadron-$\phi$(1020) Correlations in STAR}

In order to plot the $\Delta\phi$ distribution between the hadron
trigger particle (p$_{T}$~$>$~4~GeV/c) and the associated
$\phi$(1020) meson ($\langle p_{\rm T} \rangle \sim 0.9$ GeV/c), we
use $\phi$(1020) mesons identified via an invariant mass cut (1019.4
$\pm$ 7 MeV/c$^{2}$) on the decay kaon pair. The $\phi$(1020)
signal/background is only 2.2\%. In order to subtract the
background, the same angular $\Delta \phi$ correlations are
generated from a mixed event sample and the two histograms are
subtracted and normalized using the zero yield at minimum (ZYAM)
\cite{zyam}.

\begin{figure}[h!]
\centering
\includegraphics[width=0.65\textwidth]{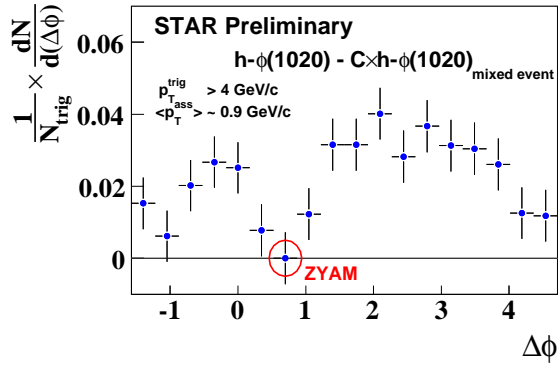}

\caption{STAR, angular correlation of hadron-$\phi$(1020) resonance.
Hadron trigger p$_{\rm T}$ $>$ 4~GeV/c and associated $\phi$(1020)
$\langle p_{\rm T} \rangle \sim 0.9$ GeV/c.}
\label{resocorrstarreso}
\end{figure}

\begin{figure}[h!]
\centering
\includegraphics[width=0.65\textwidth]{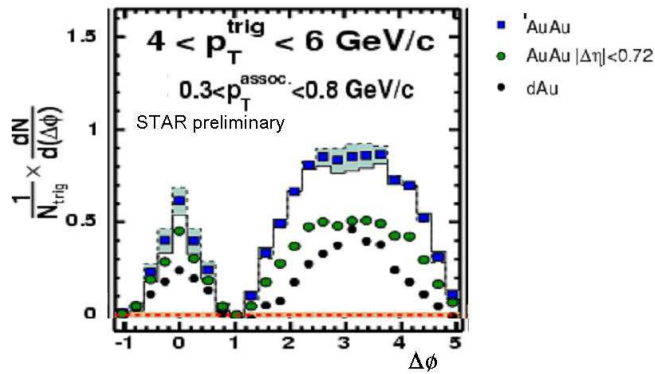}
\caption{STAR, angular correlation of hadron-hadron for 0-12\%
cental Au+Au and d+Au collisions. Hadron trigger 4~GeV/c $<$ p$_{\rm
T}$ $<$ 6~GeV/c and associated hadron 0.3~GeV/c $<$ p$_{\rm T}$ $<$
0.8~GeV/c \cite{horner06}. }
\label{resocorrstarhadrons}
\end{figure}

Figure~\ref{resocorrstarreso} shows the background subtracted
hadron-$\phi$(1020) correlation. This preliminary result is not
corrected for elliptic flow (v$_{2}$) contribution and has no
systematical error estimation. However the trend of a larger
resonance production in the away-side of the $\Delta \phi$
correlation compared to the same side is present as it is in the
angle dependent mass distribution in the previous chapter, which
might be due to energy conservation (trigger bias). Similar results
were shown previously for the charged hadron-hadron correlations by
STAR (see Figure~\ref{resocorrstarhadrons}) \cite{horner06}. In
order to study the hadron-resonance correlation in more detail, the
TOF upgrade for particle identification in STAR will improve and
extend the signal/background for resonances out to higher momentum.
A further refinement to this analysis is to reconstruct full jets
using the Calorimeter (EMC) in STAR instead of selecting a single
trigger particle.

\section{Conclusions}
Hadronic decays of resonances with different lifetimes are used to
extract information about the time evolution and temperature of the
expanding hadronic medium. To derive any additional information
about the early partonic stage of a heavy-ion collision, resonances
and decay particles need to be unaffected by the hadronic medium.
These proceedings describe a first attempt to select resonances from
jets, through angular correlation of a trigger hadron and an
associated resonance particle. Due to the low momenta of the
identified decay particles only thermally produced resonances are
presently reconstructed. Further studies on selection criteria will
be done to select higher momentum resonances and the possibility of
a full jet reconstruction with the STAR detectors will be explored.
Further theoretical studies of formation time are needed to extract
the momentum range of resonances which are formed in, and modified
by, the early partonic medium.


\begin{thebibliography}{}

\bibitem{nig05} J. Adams {\it et al.}, Phys. Rev. {\bf C71}, 044906 (2005).
\bibitem{resostar} B.I. Abelev {\it et al.}, (STAR collaboration), Phys. Rev. Lett. 97,
132301 (2006)
%nucl-ex/0604019.
\bibitem{urqmd} M. Bleicher {\it et al.},  Phys. Lett. {\bf  B530}, 81 (2002) and private communication.
\bibitem{tor01} G. Torrieri {\it et al.}, Phys. Lett. {\bf B509}, 239 (2001).
\bibitem{raf01} J. Rafelski {\it et al.}, Phys. Rev. {\bf C64}, 054907 (2001).
\bibitem{raf02} J. Rafelski {\it et al.}, Phys. Rev. {\bf C65}, 069902 (2002).
\bibitem{mar02} C. Markert {\it et al.},  hep-ph/0206260.
\bibitem{tof} B. Bonner {\it et al.}, Nucl. Instrum. Methods {\bf A508}, 181
(2003); M. Shao {\it et al.}  Nucl. Instrum. Methods {\bf A492}, 344
(2002).
\bibitem{horner06} M. J. Horner {\it et al.}, (STAR collaboration)
QM2006, nucl-ex/0701069.

\bibitem{zyam} N.N. Ajitanand {\it et al.}, Phys. Rev. {\bf C72}, 011902 (2005).
\bibitem{falter} K. Gallmeister, T. Falter, Phys. Lett. {\bf B630},
40 (2005)
% nucl-th/0502015.
\bibitem{vitev} A. Adil, I. Vitev, hep-ph/0611109.



\end{thebibliography}
\end{document}